\documentclass{natureprintstyle}

\usepackage{hyperref}
\usepackage{graphicx}
\usepackage{amssymb}
\usepackage{amsmath}

\title{Early turbulent mixing as the origin of chemical homogeneity in open star clusters}

\author{Yi Feng$^1$ and Mark R.~Krumholz$^1$}

\begin{document}

\maketitle

\begin{affiliations}
\item Department of Astronomy, University of California, Santa Cruz, CA 95064
\end{affiliations}

\begin{abstract}
The abundances of elements in stars are a critical clue to their origins. Observed star-to-star variations in logarithmic abundance within an open cluster are typically only $\sim 0.01-0.05$ over many elements\cite{2007AJ....133..694D, 2007AJ....133.1161D, pancino10a, bubar10a, de-silva11a, ting12a, reddy12b, de-silva13a, reddy13a}, significantly smaller than the variation of $\sim 0.06-0.3$ seen in the interstellar medium from which the stars form\cite{2008ApJ...675.1213R, 2012ApJ...758..133S, 2013ApJ...775..128B, 2011ApJ...730..129B, 2013ApJ...766...17L}. It is unknown why clusters are so homogenous, and whether homogeneity should also prevail in regions of lower star formation efficiency that do not produce bound clusters. Here we report adaptive mesh simulations using passively-advected scalars in order to trace the mixing of chemical elements as star-forming clouds form and collapse. We show that turbulent mixing during cloud assembly naturally produces a stellar abundance scatter at least $\sim 5$ times smaller than that in the gas, sufficient to fully explain the observed chemical homogeneity of stars. Moreover, mixing occurs very early, so that regions with efficiencies $\varepsilon \sim 10\%$ are nearly as well-mixed as those with $\varepsilon \sim 50\%$. This implies that even regions that do not form bound clusters are likely to be well-mixed, and enhances the prospects for using chemical tagging to reconstruct dissolved star clusters via their unique chemical signatures.
\end{abstract}

The question of how star clusters become chemically well-mixed has received fairly little attention. With a few exceptions\cite{carroll-nellenback13a}, work to date has been limited to simple analytic estimates\cite{murray90a}, or to calculations omitting star formation and self-gravity\cite{de-avillez02a, 2012ApJ...758...48Y}. To improve this situation, we have performed a series of simulations of star cluster formation including hydrodynamics, gravity, and optically thin radiative heating and cooling. Our simulations use the \textsc{orion} code\cite{1998ApJ...495..821T, Klein99a, 2004ApJ...611..399K}, with a new implementation of particle-mesh gravity (Methods, Figures \ref{fig:orbit}, \ref{fig:bondi}). We use initial conditions based on the ``colliding flow" model\cite{2007ApJ...657..870V}. We consider a region containing gas of number density $n_0=1 \ \rm{cm^{-3}}$ (mass density $\rho_0 = 2.1 \times 10^{-24}\ \rm{g \ cm^{-3}}$) with initial temperature $T=5000\ \rm{K}$. The gas has a random turbulent velocity $v_{\rm rms}$, for which we consider two values: $0.17$ and $1.7$ km s$^{-1}$; we refer to runs with these values as S and L, for small and large turbulence, respectively. On top of the turbulent velocity, we set up two cylindrical regions 32 pc long and 32 pc in radius, centered on the $x$-axis, with their closer ends separated by 64 pc, within which the gas has a uniform velocity $v_0=9.2$ km s$^{-1}$, directed toward the other cylinder. To trace chemical mixing the simulation includes two passive scalars $Q_L$ and $Q_R$, which have initial abundances of 1 within the left and right cylinders, respectively, and 0 elsewhere. In addition to these two simulations with smooth initial conditions, we also run a simulation C (for clumpy)\cite{carroll-nellenback13a}, in which we randomly add cold clumps one coarse cell in radius with a filling fraction of 0.05 and a number density $n_c=132.5 \ \rm{cm^{-3}}$; at this density the equilibrium temperature is such that the pressure is in equilibrium with that of the warm low-density background. Full details on the simulations are given in the Methods section.

\begin{figure*}
\centering
\includegraphics[width=\textwidth]{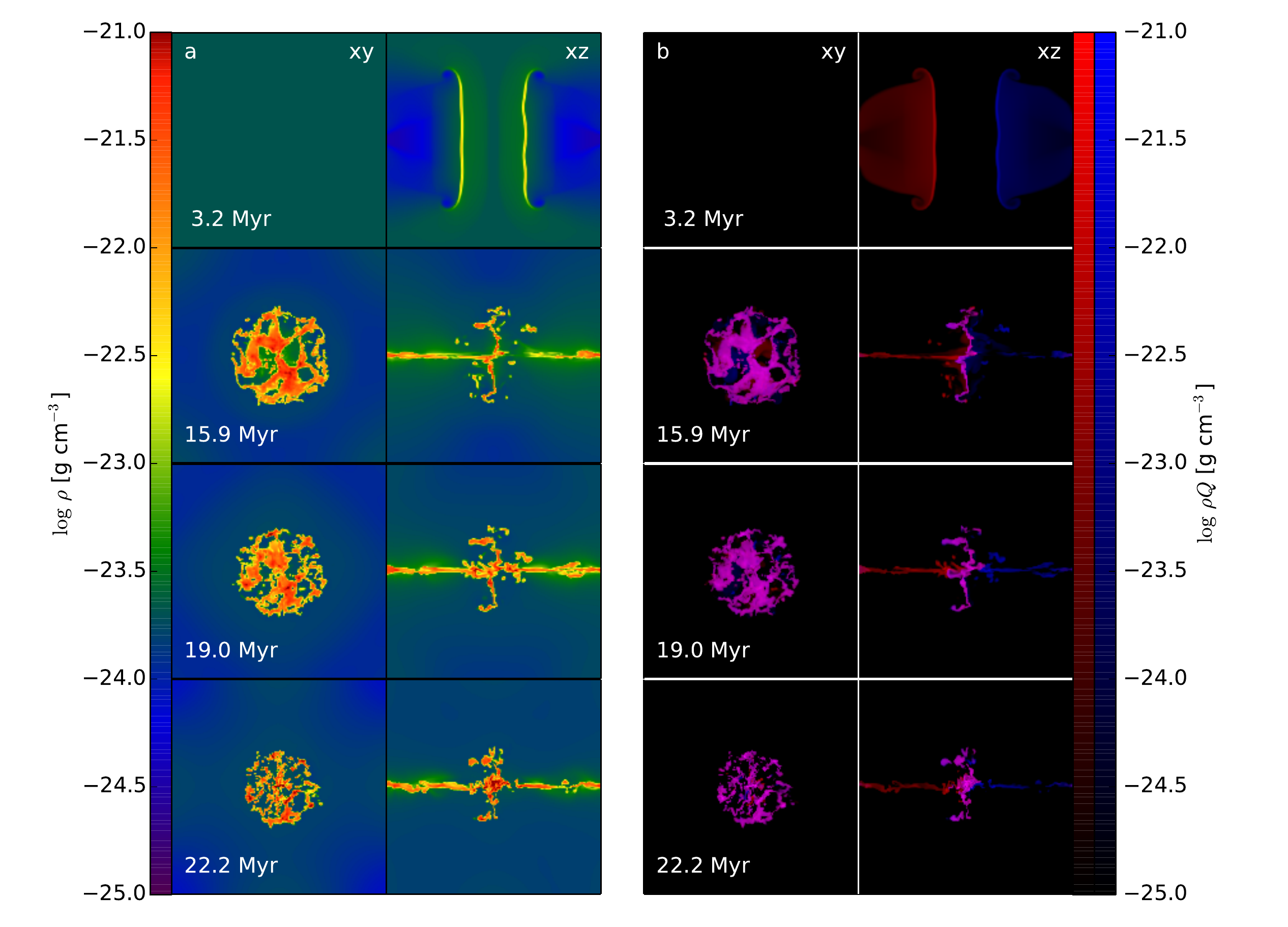}
\caption{
\label{fig:slices}
Slices through simulation S at a variety of times, showing the total density and the densities of the passive scalar fields. (a) Gas density $\rho$ in run S on slices in the $xy$ (left column) and $yz$ (right column) planes. The rows show increasing times in the simulation, as indicated in each row. (b) Density of passive scalars $\rho Q_L$ (red) and $\rho Q_R$ (blue) at the same times and in the same planes as in panel (a). The densities of the two tracers have been mapped to the red and blue channels of the image, so that cells containing equal contributions from the two streams appear as purple, with the intensity of the purple color proportional to the logarithm of the total density. In contrast, cells dominated by one passive scalar or the other appear as red or blue in color.
}
\end{figure*}

\begin{figure}
\begin{center}
\includegraphics[width=0.5\textwidth]{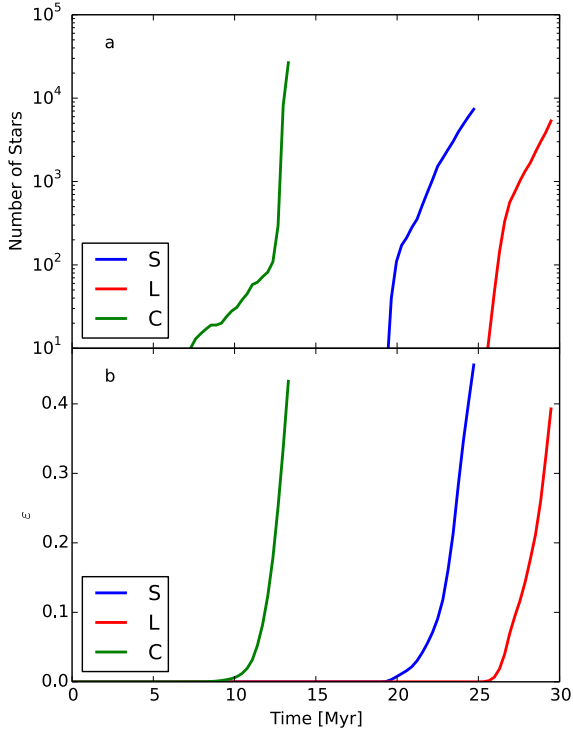} 
\end{center}
\caption{
\label{fig:nstareps}
Number of stars and star formation efficiency as a function of time. (a) Number of stars in simulations S, L, and C. (b) Star formation efficiency $\varepsilon$ versus time in the same simulations.
}
\end{figure}

The overall evolution of our simulations is very similar to previous colliding flow simulations\cite{2007ApJ...657..870V,2008ApJ...674..316H, carroll-nellenback13a} (Figure \ref{fig:slices}). The two streams of gas converge rapidly, and compression of material at the heads of the two cylinders leads to thermal instability and the formation of a cold phase even before the two flows collide. For runs S and L, the two streams collide just before $10$ Myr of evolution, and this produces a dense, cold, turbulent layer that is gravitationally unstable. The layer begins to form stars at $\approx 19$ Myr in run S, and approximately 50\% of the gas in the two streams has been converted to stars by $\approx 25$ Myr of evolution (Figure \ref{fig:nstareps}). In run L star formation begins at $\approx 25$ Myr, and follows a similar time evolution to run S thereafter. In run C the two streams collide at $\approx 6$ Myr because the dense cold phase is less decelerated by the warm phase. Star formation begins immediately after collision, and $\sim 40\%$ of the gas in the two streams has been converted to stars by $\approx 13$ Myr of evolution. We note that all our simulations have star formation rates that exceed observationally-inferred values\cite{tan06a, krumholz07e}, but we select this scenario to examine precisely because its rapidity minimizes the time available to fully mix out chemical inhomogeneities. Our results should therefore represent lower limits on the true amount of mixing.

\begin{figure*}
\centering
\includegraphics[width=0.8\textwidth]{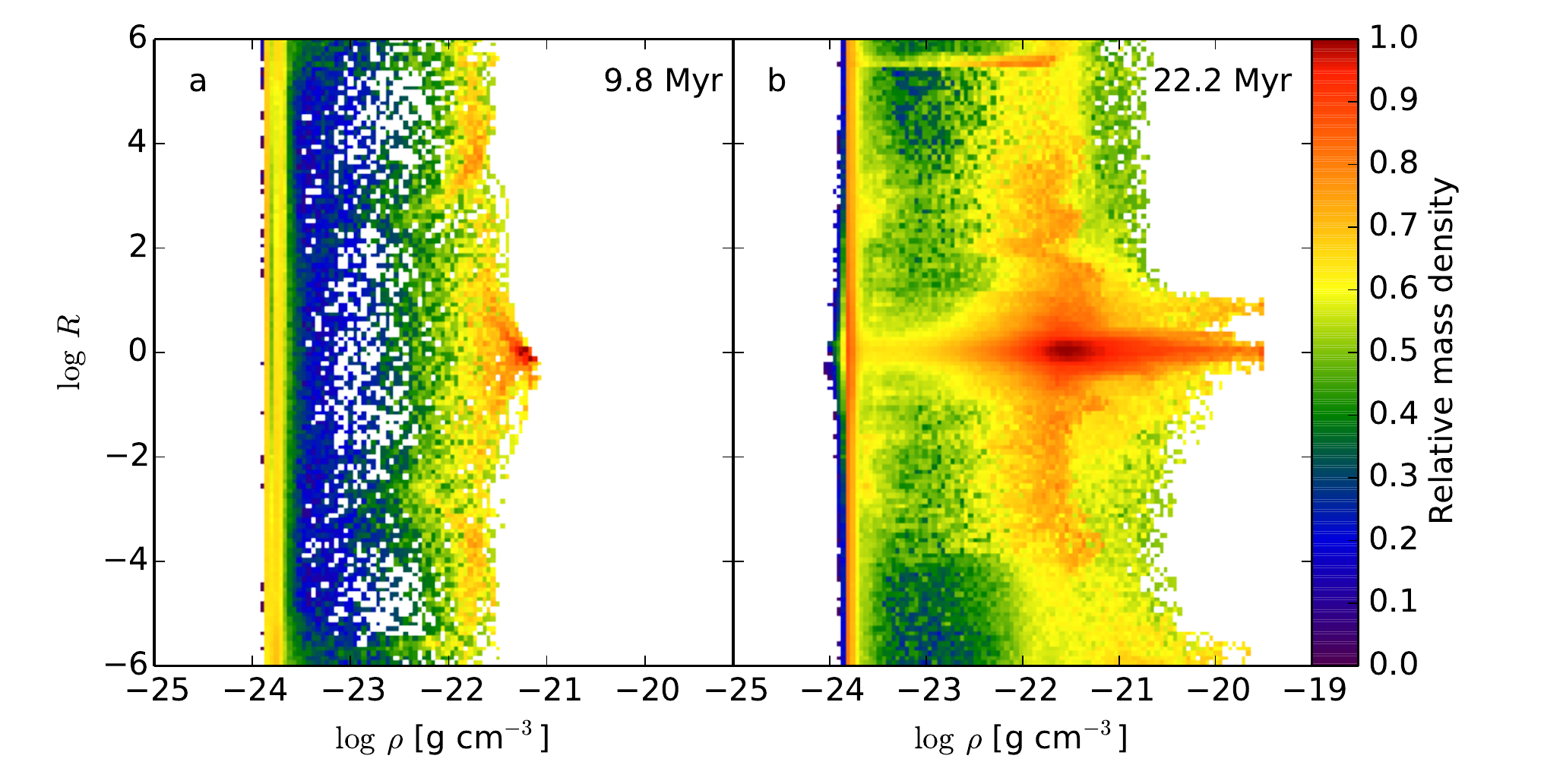} 
\caption{
\label{fig:tracerratios}
Distribution of gas in simulation S in density and mixing ratio at two different times. The color in each 2D pixel indicates the relative fraction of mass in the corresponding bin of $(\rho, R)$, where $R = Q_L/Q_R$ is the ratio of the two passive scalars. Panel (a) shows the result at $t=9.8$ Myr, just as the two streams are beginning to collide, and panel (b) shows the result at $t=22.2$ Myr, just after the onset of rapid star formation. Note that some of the features seen in panel (b), including the streaks near $R\approx 3\times 10^{-6}$ and $R \approx 10^{-1}$, are transients due to the chaotic nature of the mixing process. Similar features appear at other times and for different simulation resolutions, but they come and go essentially randomly. Only the must more prominent structure near $R=0$ is persistent.
}
\end{figure*}

By the onset of star formation, the interaction region where the flows have collided is reasonably well-mixed by the turbulence (Figure \ref{fig:slices}). The ratio of passive scalar concentrations $R=Q_L/Q_R$ is very broad for material at densities up to $\approx 10^{-22}$ g cm$^{-3}$, reflecting the broad range of abundances in low-density gas. However, in gas with densities $\sim 10^{-21}$ g cm$^{-3}$ or higher, the range of compositions is dramatically reduced (Figure \ref{fig:tracerratios}). For the densest gas, the full range in $R$ is at most a decade, and the vast majority of the mass is spread over an even smaller range. This densest gas is produced in regions where the two flows are converging and mixing efficiently, and it is these regions that produce stars.

\begin{figure}
\begin{center}
\includegraphics[width=0.5\textwidth]{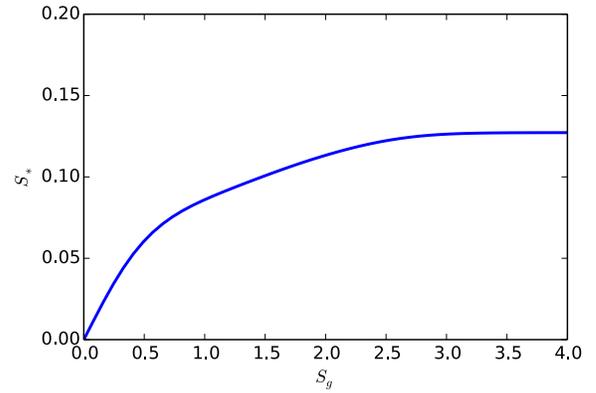}
\end{center}
\caption{
\label{fig:sstarsg}
Stellar abundance scatter $S_*$ versus gaseous abundance scatter $S_g$ for the final time in simulation S.
}
\end{figure}

To assess how this affects stellar abundances, note that the abundance scatter for some element in a collection of stars is formally defined as $S_* = [\sum (\log a_{*,i} - \overline{\log a_*})^2/N]^{1/2}$, where the sum runs over all $N$ stars present at any time, $a_{*,i}$ is the abundance of star $i$, and $\overline{\log a_*} = (1/N) \sum_i \log a_{*,i}$ is the mean logarithmic abundance of the stars. To compute this quantity from our simulations, let $a_L$ and $a_R$ be the abundances of some element of interest in the left and right streams, respectively. Without loss of generality we can choose $a_L<a_R$. For each star $i$ formed in the simulations, we know the masses $M_{L,i}$ and $M_{R,i}$ contributed by each stream. The abundance $a_{*,i}$ of that star is therefore $a_{*,i} = (a_L M_{L,i} + a_R M_{R,i}) / (M_{L,i}+M_{R,i})$. Note that the actual values of $a_L$ and $a_R$ need not be chosen before the simulations are run, since the only quantities actually measured from the simulations are $M_{L,i}$ and $M_{R,i}$. We can therefore use a single simulation to compute $S_*$ for an arbitrary value of the gas abundance ratio $a_R/a_L$ (see Figure \ref{fig:sstarsg}). 

We show in the Methods section that the dependence of $S_*$ on $a_R/a_L$ is characterized by two limiting cases: when $a_R/a_L \approx 1$, we have $S_* \approx 2 \sigma_Y S_g \equiv S_{\rm slope} S_g$. Here $Y_i = M_{R,i}/(M_{L,i}+M_{R,i})$ is the mass fraction in star $i$ provided by the right stream, $\sigma_Y = \sqrt{\sum (Y_i-\overline{Y_i})^2/N}$ is the dispersion of the $Y_i$ values, and we have defined the gas scatter as $S_g = \{[(\log a_L - \overline{\log a_g})^2 + (\log a_R - \overline{\log a_g})^2]/2\}^{1/2}$, where $\overline{\log a_g} = (\log a_L + \log a_R)/2$ is the mean logarithmic abundance in the two streams. In the opposite limit, when $a_R/a_L \gg 1$, we have $S_* \approx \sigma_{\log Y} \equiv S_{\rm limit}$, where $\sigma_{\log Y}$ is the dispersion of $\log Y_i$. Intuitively, the reduction in abundance scatter is at its minimum when the gas is close to homogenous already, and $S_{\rm slope}$ characterizes the factor by which the gas abundance scatter is reduced in this limit. The quantity $S_{\rm limit}$ is the maximum possible stellar abundance scatter no matter how inhomogeneous the gas is. 

\begin{figure}
\centering
\includegraphics[width=0.5\textwidth]{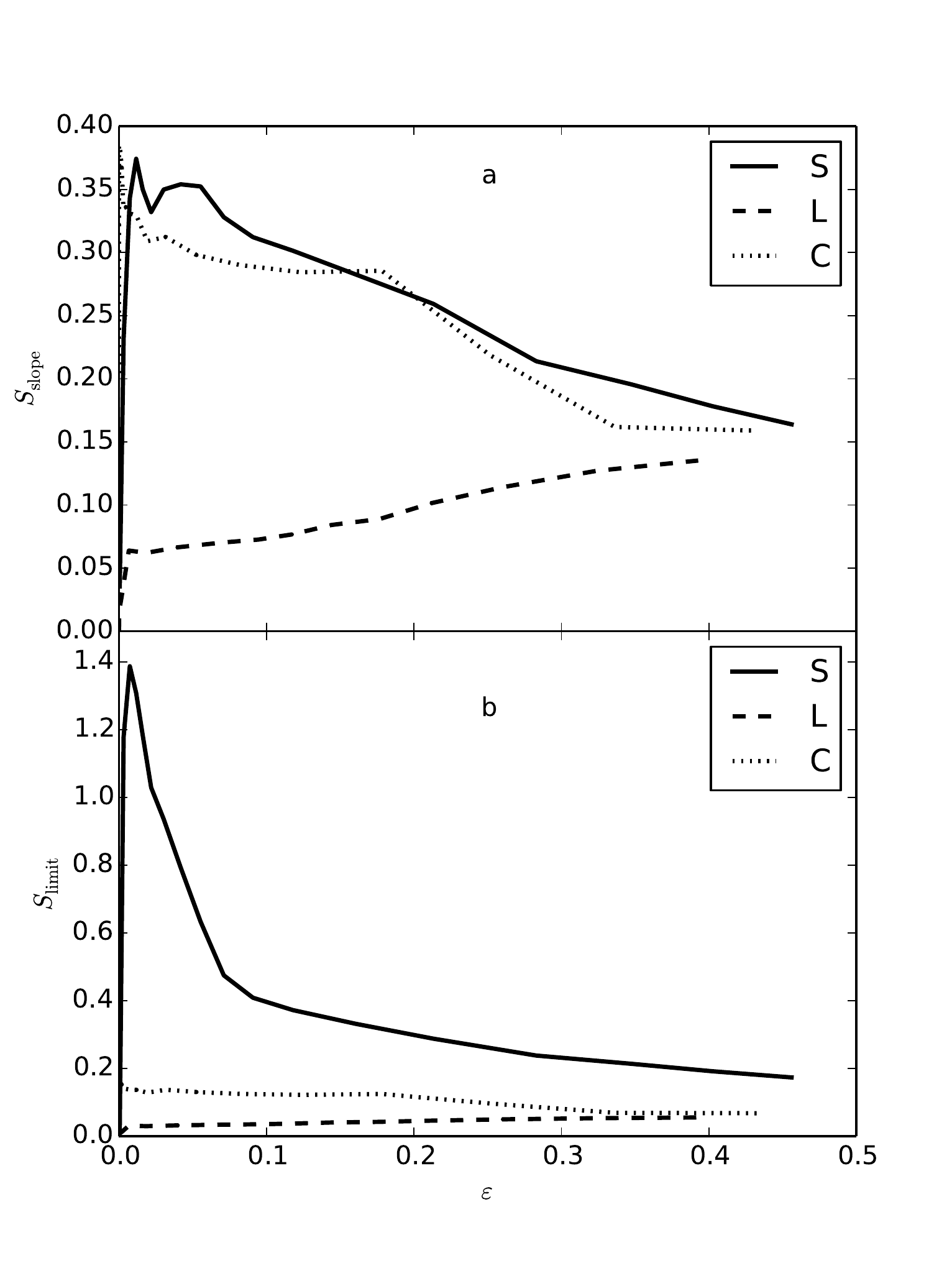} 
\caption{
Two measures of the stellar abundance scatter as a function of star formation efficiency in simulations S, L, and C. (a) $S_{\rm slope}$ versus star formation efficiency $\varepsilon$, where $S_{\rm slope}$ indicates the factor by which the gaseous abundance scatter $S_g$ is reduced by star formation in the limit where $S_g \ll 1$. (b) $S_{\rm limit}$ versus star formation efficiency $\varepsilon$, where $S_{\rm limit}$ is the maximum possible stellar abundance scatter in the gas where $S_g \gg 1$.
\label{fig:svseps}
}
\end{figure}

We define the star formation efficiency $\varepsilon = M_*/2M_{\rm inf}$, where $M_*$ is the total stellar mass and $M_{\rm inf}$ is the mass in one of the stream; $M_{\rm inf} = 6.5\times 10^3$ $M_\odot$ in runs S and L, and $4.9\times 10^4$ $M_\odot$ in run C. The general evolution of both $S_{\rm limit}$ and $S_{\rm slope}$ with $\varepsilon$ in run S is a rapid rise from 0 as the first stars form (Figure \ref{fig:svseps}), followed by a rapid fall by the time $\varepsilon$ reaches $\sim 0.02$. At values of $\varepsilon > 0.1$, we have $S_{\rm limit} \lesssim 0.4$ and $S_{\rm slope} \lesssim 0.3$, indicating that a relatively small abundance inhomogeneity will be reduced by a factor of at least 3 in the star formation process, and that even a very large inhomogeneity will produce at most $\sim 0.4$ dex of scatter in the resulting stars. By the time the star formation efficiency reaches $\sim 30\%$, the reduction in scatter is close to a factor of 5, and the absolute upper limit on the scatter is $\sim 0.2$ dex. In run L, the stronger turbulence delays the onset of star formation and allows more rapid mixing at early times, so the stellar scatter starts small and very gradually increases with time. However, it is always smaller than at the corresponding value of $\varepsilon$ in run S. Similarly, $S_{\rm limit}$ is smaller in run C than in run S, likely due to the stronger global collapse in the clumpy run\cite{carroll-nellenback13a}. However, $S_{\rm slope}$ is nearly identical in runs S and C. This suggests that clumpiness does not significantly alter the amount of mixing much where $S_g \ll 1$. We have also conducted convergence studies to verify that our results for mixing are robust against changes in numerical resolution (Methods, Figures \ref{fig:converge} and \ref{fig:sslope_limit_converge}).

Figure \ref{fig:svseps} indicates that the process of star formation leads to a great deal of chemical homogenization as soon as even very modest star formation efficiencies are achieved. For realistic efficiencies, which are probably in range of $\sim 10-50\%$ on the scale of star clusters\cite{krumholz14c}, we should expect the abundance scatter to be reduced by at least a factor of $\sim 4-6$ compared to that in the gas from which the stars are formed, and even in the most chemically inhomogeneous environments the scatter will be no more than a few tenths of a dex. Since observed gas abundance scatters are $S_g \sim 0.06 - 0.3$ dex over size scales of $\sim 0.1-1$ kpc\cite{2008ApJ...675.1213R, 2011ApJ...730..129B, 2012ApJ...758..133S, 2013ApJ...775..128B, 2013ApJ...766...17L}, a factor of $\sim 5$ reduction in the stellar abundance scatter compared to this is sufficient to fully explain the observed scatter $S_*\sim 0.01-0.05$ dex seen in open clusters and moving groups.

Moreover, our results are also very encouraging for the prospects of chemical tagging as a method of reconstructing the star formation history of the Milky Way, and identifying potential ``Solar siblings", stars born in the same cluster as the Sun\cite{portegies-zwart09a, bland-hawthorn10a}. We find that both $S_{\rm limit}$ and $S_{\rm slope}$ reach values $\sim 0.1-0.3$ even at low star formation efficiencies $\sim 0.1$, and that the degree of mixing increases only modestly as $\varepsilon$ rises from $\sim 0.1$ to $\sim 0.5$. Since star formation sites with $\varepsilon \sim 0.1-0.3$ are likely the progenitors of the majority of field stars, while those with $\varepsilon\sim 0.5$ likely represent the sites of bound cluster formation, our results imply that the clusters and moving groups that have been studied for chemical homogeneity thus far are \textit{not} atypical in their degree of chemical mixing. They are at most marginally better mixed. Thus it is likely that even those stars that did not form in bound clusters will be chemically similar to their neighbors formed at the same point in space and time, and that these unique chemical signatures can serve as a fingerprint to identify these common formation sites even as stars disperse throughout the Galaxy. Indeed, with a group finding technique\cite{2013MNRAS.428.2321M} and high resolution data, recent work\cite{2014MNRAS.438.2753M} shows evidence that chemical tagging of field stars does identify coeval groups of stars. We discuss the implications and broader context of our work in more detail in the Methods section.

\begin{addendum}
\item[Acknowledgements]This work was funded by NSF CAREER grant AST-0955300, NASA ATP grant NNX13AB84G, NASA TCAN grant NNX14AB52G, and NASA through Hubble Award Number 13256 issued by the Space Telescope Science Institute, which is operated by the Association of Universities for Research in Astronomy, Inc., under NASA contract NAS 5-26555. The simulations for this research were carried out on the UCSC supercomputer Hyades, which is supported by the NSF (award number AST-1229745).
\item[Author Contributions]Y.~Feng ran the simulations, produced all the figures, and wrote parts of the text. M.~R.~Krumholz aided in the interpretation and wrote other parts of the text.
\item[Author Information]Reprints and Permissions information
is available at \url{npg.nature.com/reprintsandpermissions}. Authors declare they have no competing financial interests. Correspondence and requests for materials should be addressed to M.~R.~Krumholz (mkrumhol@ucsc.edu).
\end{addendum}

\clearpage

\section{Methods.\\\\}

\noindent
\textbf{Equations and Algorithms.}\\

We perform simulations using the parallel adaptive mesh refinement (AMR) code \textsc{orion}. \textsc{Orion} utilizes a conservative second order Godunov scheme to solve the equations of compressible gas dynamics coupled to a multi-grid method to solve the Poisson equation for gas self-gravity\cite{1998ApJ...495..821T, Klein99a}. We treat radiative heating and cooling by parameterized heating and cooling curves, which we take from the approximation of Koyama \& Inutsuka\cite{2002ApJ...564L..97K}. In self-gravitating collapse problems, it is necessary to cut off the collapse at finite resolution in order to render the problem computationally tractable, and \textsc{orion} handles this problem by replacing regions that become Jeans-unstable at the finest allowed resolution by sink particles\cite{2004ApJ...611..399K}. The full set of equations solved by the code is
\begin{eqnarray}
\frac{\partial \rho}{\partial t}+\nabla \cdot(\rho \mathbf{v})& = & -\sum_i \dot{M_i} W(\mathbf{x}-\mathbf{x}_i) \\
\frac{\partial (\rho \mathbf{v})}{\partial t}+\nabla \cdot(\rho \mathbf{v}\mathbf{v})& =& \nabla P -\rho \nabla \phi \nonumber \\
& & {} - \sum_i \dot{\mathbf{p}_i} W(\mathbf{x}-\mathbf{x}_i) \\
\frac{\partial (\rho e)}{\partial t}+\nabla \cdot[(\rho e+P) \mathbf{v}] & = & \rho \mathbf{v} \nabla{\phi} - \sum_i \dot{\varepsilon}_i W(\mathbf{x}-\mathbf{x}_i)
\nonumber \\
& & {}
-n^2 \Lambda + n \Gamma \\
\frac{\partial (\rho Q_{k})}{\partial t}+\nabla \cdot(\rho Q_{k} \mathbf{v})& = & -\sum_i \dot{M}_{k,i} W(\mathbf{x}-\mathbf{x}_i)\\
\nabla^2 \phi & = & 4 \pi G \rho  + 4\pi G \sum_i M_i \delta(\mathbf{x}-\mathbf{x}_i)\\
\frac{d}{dt} M_i & = &\dot{M_i}\\
\frac{d}{dt} M_{k,i} & = & \dot{M_{k,i}} \\
\frac{d}{dt} \mathbf{x}_i& = &\frac{\mathbf{p}_i}{M_i}\\
\frac{d}{dt} \mathbf{p}_i & = &-M_i \nabla{\phi} + \dot{\mathbf{p}_i}
\end{eqnarray}
where $\rho$, $P$, and $\mathbf{v}$ are the fluid density, pressure, and velocity, respectively, $e = (1/2)v^2 + P/[\rho (\gamma -1)]$ is the specific energy of the gas, and $\phi$ is the gravitational potential. We do not attempt to model the transition from atomic to molecular gas, and thus we adopt and constant ratio of specific heats $\gamma=5/3$, as appropriate for a monatomic ideal gas. Passive scalars or ``colors'' are denoted with $Q_{k}$, where $k$ is the index of the tracer in question. We use these to represent and track abundance patterns in the gas.

Terms subscripted by $i$ refer to sink particles, which represent stars; $\mathbf{x}_i$, $M_i$, $\mathbf{p}_i$, and $M_{k,i}$ are the position, mass, momentum, and mass of passive scalar $k$ in the $i$th star, and $\dot{M}_i$, $\dot{\mathbf{p}}_i$, $\dot{\varepsilon}_i$, and $\dot{M}_{k,i}$ are the rates at which those stars add or remove mass, momentum, energy, and the mass of the $k$th tracer from the gas. The quantity $W_i$ is the weighting kernel that spreads the stellar interaction over some number of computational cells. These quantities are all computed following the sink particle algorithm introduced by Krumholz et al.\cite{2004ApJ...611..399K}, which estimates the accretion rates onto sink particles by fitting the gas around them to Bondi-Hoyle flow.

The quantities $n$, $\Gamma$, and $\Lambda$ are the number density, heating function, and cooling function. Since we are interested in simulating flows in the atomic interstellar medium that lead to the formation of star clusters, we adopt a mean molecular weight of 1.27, so $n = \rho/(1.27m_{\rm H})$, and we adopt the approximate heating and cooling functions suggested by Koyama \& Inutsuka\cite{2002ApJ...564L..97K}, 
\begin{eqnarray}
\label{eq:heatfunction}
\Gamma & = & 2.0 \times 10^{-26} \mbox{ erg s}^{-1} \\
\frac{\Lambda}{\Gamma} & = & 10^7 \exp\left(\frac{-1.184 \times 10^5}{T + 1000}\right) 
\nonumber\\
& & {} + 1.4 \times 10^{-2} \sqrt{T} \exp \left(\frac{-92}{T}\right) \ \rm {cm^3}
\label{eq:coolfunction}
\end{eqnarray}
Here the temperature $T$ is in K, and is given by $T = (e-v^2/2)/[(\gamma-1) k_B]$. Physically, $\Gamma$ represents the rate of photoelectric heating per particle, while $\Lambda$ describes cooling due to emission in the Lyman $\alpha$ and C$^+$ 158 $\mu$m lines, which dominate cooling at high and low temperatures, respectively.

We use the AMR capability in our code to increase the resolution in regions undergoing gravitational collapse. We refine by a factor of 2 any cells in which the local density exceeds the Jeans density
\begin{equation}
\label{eq:rhoj}
\rho_J = J^2 \frac{\pi k_B T}{\mu m_{\rm H} G \Delta x^2},
\end{equation}
where we use a Jeans number $J = 1/8$, $\mu = 1.27$ is the mean molecular weight, and $\Delta x$ is the cell size. Refinement continues up to some specified maximum level. If the density exceeds the Jeans density on this maximum level, evaluated with a Jeans number $J=1/4$, we introduce a sink particle.\\

\noindent
\textbf{Sink Particle Algorithm.}\\

For the purposes of this computation, we have modified the implementation of sink particles in the \textsc{orion} code slightly from the method described by Krumholz et al.\cite{2004ApJ...611..399K}. First, in addition to tracking the mass and momentum of sink particles as in the original method, we also track the masses of passive scalars. We compute the rates at which passive scalars are incorporated into sink particles by assuming that the accretion rate for each passive scalar in a given computational cell is equal to the overall mass accretion rate from that cell multiplied by the concentration $Q_k$ of the passive tracer in that cell. Thus the total mass of passive scalar $M_k = \int \rho Q_k\, dV + \sum_i M_{k,i}$ over the entire computational grid plus that in sink particles is conserved by the accretion process, as are the concentrations $Q_k$ in the cells from which accretion occurs. Since we initialize the passive scalar abundances $Q_L$ to unity in the left stream, and zero elsewhere, the mass contributed to star $i$ from stream $L$ is identical to the mass $M_{L,i}$ of the passive scalar in that star, and similarly for $M_{R,i}$.

Second, in the original Krumholz et al.\cite{2004ApJ...611..399K} method, the velocities of sink particles were updated by calculating the gravitational force between every cell and every sink particle. The code then performed an operator-split step during which the sink particle positions and velocities were evolved under their mutual gravitational interaction using a sub-cycled ordinary differential equation solver. While this approach is highly accurate, and allows the code to correctly evolve sink particle orbits even when they are smaller than the size of a hydrodynamic cell, the computational cost of this method scales as the number of sink particles times the number of computational cells, plus the square of the number of sink particles. This is prohibitively expensive for the large number of sink particles ($\gtrsim 5000$) that form in the simulations we present here.

For this reason, we have implemented a particle-mesh (PM) method to update sink particle positions and velocities. Before solving the Poisson equation, we assign the mass carried by sink particles to the computational grid, so that this mass is included when solving for the gravitational potential. We perform this mass assignment using a cloud-in-cell interpolation (CIC) scheme\cite{hockney88a}. Once we have obtained the potential, we update the positions and velocities of the particles. For the velocity update, we compute the accelerations of the particles from the gradient of the potential returned by the Poisson solve, interpolated in space to the positions of the particles using the same CIC interpolation scheme. The use of the same interpolation for the mass assignment and force computation steps ensures that self-forces vanish to the accuracy of the Poisson equation solution.

We have performed two tests of this implementation. The first is maintaining the orbit of a binary system. We place two sink particles of mass $10 \ \rm M_{\odot}$ into a computational domain that runs from $[-2.5 \times 10^{13}, 2.5 \times 10^{13}]$ cm in each direction. Particle 1 is initially placed at $(3.125 \times 10^{12},0)$ cm with velocity $(0,1.033 \times 10^{7})$ cm s$^{-1}$. Particle 2 is initially placed at $(-3.125 \times 10^{12},0)$ cm with velocity $(0,-1.033 \times 10^{7})$ cm s$^{-1}$. The separation and velocity we have chosen are such that the particles should perform a circular orbit at constant radius centered on the origin. In addition to the two sink particles, we fill the computational domain with a uniform, isothermal gas of density $1.0 \times 10^{-22} \rm \ g \ cm^{-3} $ and sound speed $1.3 \times 10^7 \rm \ cm \ s^{-1}$. The density and sound speed are such that the mass accreted onto the particles per orbit should be a negligible fraction of their initial mass, and thus interaction with the gas should have no effect on the orbit.

\begin{figure}
\begin{center}
\includegraphics[width=0.5\textwidth]{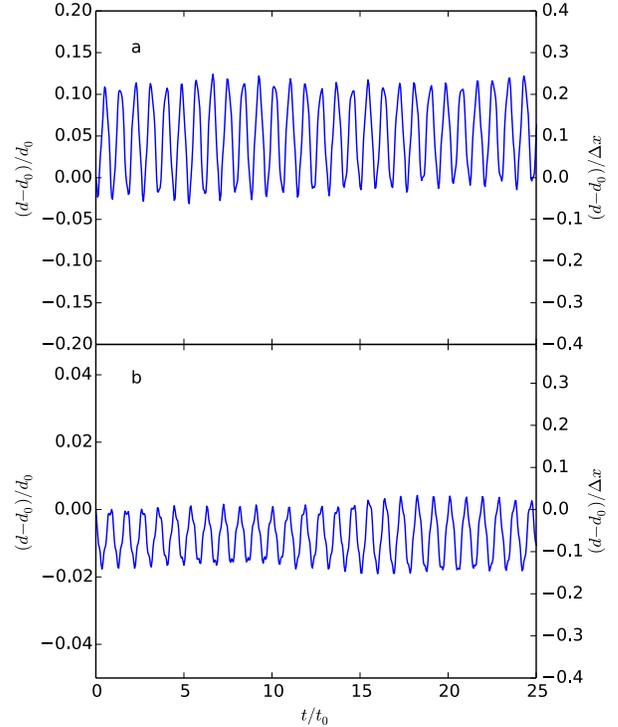}
\end{center}
\caption{
\label{fig:orbit}
Variation in distance $d$ between two stars in a test of how well our new particle-mesh gravity implementation can maintaing the orbit of a binary. (a) Distance between two stars $d$ minus initial distance $d_0$, in a test with $d_0=2\Delta x$, where $\Delta x$ is the cell size. The left axis shows the $d-d_0$ normalized to $d_0$, and the right axis shows it normalized to $\Delta x$. Perfect accuracy would be a flat line at $d-d_0 = 0$. (b) Same as (a), but for a test with $d_0 = 8\Delta x$, so the two stars are initially separated by 8 cells.
}
\end{figure}

We perform the test at two different resolutions: 16 and 64 cells per linear dimension, corresponding to cell sizes of $3.125 \times 10^{12}$ cm and $7.8125\times 10^{11}$ cm, respectively. Thus the particles are separated by only 2 computational cells at the lower resolution, and 8 cells at the higher resolution. In Figure \ref{fig:orbit}, we show the separation between the two sink particles tracked over many orbits. We see that the algorithm maintains the orbital separation to a precision of $\sim 10\%$ in the lower resolution test, and $\sim 1\%$ in the higher resolution test, with no apparent secular drift. In both cases the error is roughly $1/10$ the size of a computational cell.  Given that the forces felt by the particles are only interpolated to an accuracy of one cell, this is the best precision that could be expected.

The second test is Bondi accretion. We place a sink particle of mass $10 \ \rm M_{\odot}$ at the center of a computational grid that is filled with an isothermal gas of sound speed $1.3 \times 10^7$ cm s$^{-1}$, so that the Bondi radius of the particle is $7.85 \times 10^{12}$ cm. The computational grid is $1.4 \times 10^{14} \ \rm cm$ on a side, and has a linear resolution of 256 cells, so that the size of a cell is $\Delta x = r_{\rm B}/14.4$, and the length of the computational box is $L = 17.8 r_{\rm B}$. We initialize the density and velocity profile of the gas to the analytic solution for Bondi accretion, and then allow the computation to evolve for a time $t = 5 r_{\rm B}/c_s$.

\begin{figure}
\begin{center}
\includegraphics[width=0.5\textwidth]{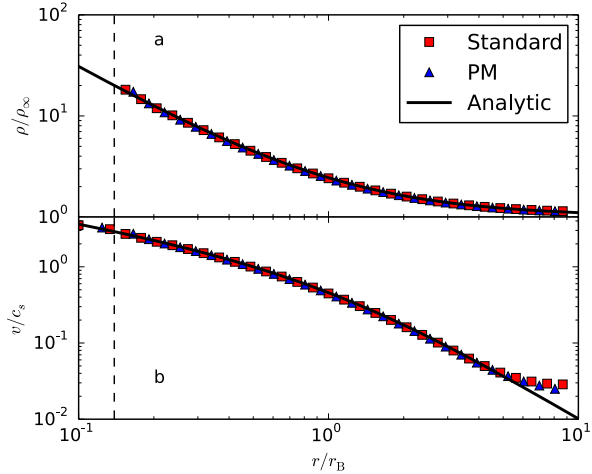}
\end{center}
\caption{
\label{fig:bondi}
Comparison between the analytic solution for Bondi accretion and the numerical results produced by an \textsc{orion} simulation. (a) Density normalized to density at infinity $\rho/\rho_\infty$ versus radius normalized to the Bondi radius $r/r_B$. We show the analytic solution (black line), the result using \textsc{orion} with its standard implementation of sink particle gravity (red squares), and the result using our newly-implemented particle-mesh gravity method. The numerical results show averages over radial bins. To prevent the numerical results from lying completely on top of on another and from obscuring the line for the exact result, we show only every fourth radial bin, and the bins we show are offset between the two simulations. The dashed vertical line shows the accretion kernel radius of 2 cells. (b) Same as (a), but now showing the infall velocity normalized to the sound speed, $v/c_s$.
}
\end{figure}

We run this test twice, once with the original \textsc{orion} sink particle implementation, and a second time with our new PM method. We show the results of both tests in Figure \ref{fig:bondi}. We can see that the algorithm maintains the density and infall velocity outside the accretion kernel quite well. The accretion rates are also close to the analytical result, with errors of 3.4\% for standard algorithm and 5\% for PM algorithm, respectively. The small differences in velocity between the two algorithms at $r/r_{\rm B} \sim 10$ are to be expected, because the gas at this distance is near the edge of the computational box, and the algorithms differ slightly in how they treat boundary conditions. The PM method imposes periodic boundary conditions on the potential, such that the gravitational force exerted by the particle goes to zero smoothly as the distance from the particle approaches half the size of the computational box. In contrast, the standard method simply uses a $1/r^2$ force law for all cells, so the force does not go to zero smoothly at the box edge.\\

\noindent
\textbf{Initial Conditions and Resolution.}\\

Runs S and L both start with a uniform density $n_0 = 1$ cm$^{-3}$ (mass density $\rho_0 = 2.1\times 10^{-24}$ g cm$^{-3}$) and a temperature of $T_0 = 5000$ K; given our choice of heating and cooling functions, this is the equilibrium temperature at that density. In addition to the uniform velocity fields imposed within the converging cylinders (see main text), we impose a turbulent velocity field with a dispersion of $0.17$ (run S) or $1.7$ (run L) km s$^{-1}$. We generate this field in Fourier space by choosing random phases and drawing amplitudes following a power spectrum that is flat at wavenumbers $k$ in the range $4\leq k L_{\rm box} \leq 8$ and zero elsewhere. Here $L_{\rm box} = 128$ pc is the size of the (cubical) computational domain. Our simulations use periodic boundary conditions, and for runs S and L we use a base grid of 256$^3$ for our coarsest level, plus 2 levels of refinement. Thus the base grid resolution is $\Delta x = 1/2$ pc, and the minimum cell size is $\Delta x_{\rm min} = 1/8$ pc.

The setup is identical for run C, except that in 5\% of the coarse cells we replace the warm medium with a cold clump with number density to $n_c = 132.5$ cm$^{-3}$ and temperature to $T_c=37.7$ K, which is the equilibrium temperature at this density. We choose this density and temperature so that $n_c T_c = n_0 T_0$, and the cold clumps are initially in pressure balance with the surrounding warm gas. We randomly choose which cells will be cold rather than warm, and the probability of a cell being cold is independent of whether it is part of one of the streams or is part of the medium between the streams.\\

\noindent
\textbf{Characterizing the Stellar Scatter.}\\

Here we show that the function $S_*(S_g)$, which characterizes the stellar abundance scatter as a function of the initial gas abundance scatter, is a linear function in the limit $S_g \rightarrow 0$, and reaches a finite limiting value as $S_g \rightarrow \infty$. 

First consider the latter case, $S_g \rightarrow\infty$. Without loss of generality we will assume $a_L < a_R$. Since we are working in the limit $S_g\rightarrow \infty$, this implies that $a_R/a_L\rightarrow \infty$ as well. We define $X_i = M_{L,i} / (M_{L,i} + M_{R,i})$ and $Y_i = M_{R,i} / (M_{L,i} + M_{R,i})$ as the mass fractions in star $i$ coming from the left and right streams, respectively, so that $a_{*,i} = X_i a_L + Y_i a_R$. As long as $Y_i \neq 0$ (i.e., as long as there is any mixing at all), then in the limit $a_R/a_L \rightarrow \infty$ we have $a_{*,i} \rightarrow Y_i a_R$. Thus the mean stellar abundance is
\begin{equation}
\overline{\log a_*} \rightarrow \frac{1}{N} \sum_i \log (Y_i a_R)
= \log a_R + \overline{\log Y},
\end{equation}
where $\overline{\log Y} = (1/N)\sum_i \log Y_i$ is the mean value of $\log Y$ over all stars. The scatter therefore approaches
\begin{eqnarray}
S_* & \rightarrow & \sqrt{\frac{1}{N} \sum_i \left[\log (Y_i a_R) - \log a_R - \overline{\log Y} \right]^2} \\
& = & \sqrt{\frac{1}{N} \sum_i (\log Y_i - \overline{\log Y})^2} \equiv \sigma_{\log Y},
\end{eqnarray}
where the quantity $\sigma_{\log Y}$ is simply the scatter in the logarithm of the mass fraction contributed by each stream. Intuitively, this makes perfect sense: if one incoming stream contains iron and the other does not, then clearly the scatter in the logarithmic iron abundance must reduce to the scatter in the logarithm of the mass fraction provided by the iron-bearing gas.

Now consider the opposite limit, $S_g \rightarrow 0$, in which case $a_L \approx a_R$. To analyze this limit, we set $a_R = (1 + 2\epsilon) a_L$ and take the limit $\epsilon \rightarrow 0$. Inserting these values into the definition of $S_g = \{[(\log a_L - \overline{\log a_g})^2 + (\log a_R - \overline{\log a_g})^2]/2\}^{1/2}$, Taylor expanding about $\epsilon = 0$, and dropping terms beyond leading order, we obtain
\begin{equation}
S_g = \frac{\epsilon}{\ln 10}.
\end{equation}
Similarly, the mean stellar abundance can be expanded to give to leading order
\begin{equation}
\overline{\log a_*} = \log a_L + \frac{2\epsilon}{\ln 10} \overline{Y},
\end{equation}
where $\overline{Y} = (1/N)\sum_i Y_i$ is the mean value of $Y_i$. The stellar abundance scatter thus becomes
\begin{equation}
S_* = \frac{2\epsilon}{\ln 10} \sqrt{\frac{1}{N}\sum_i (Y_i-\overline{Y})^2} \equiv \frac{2\epsilon}{\ln 10} \sigma_Y
\end{equation}
to leading order, where $\sigma_Y$ is the dispersion in mass fraction. We therefore have
\begin{equation}
S_* \approx 2\sigma_Y S_g
\end{equation}
in the limit $S_g \rightarrow 0$.\\

\noindent
\textbf{Convergence.}\\

Convergence is a critical issue for any calculation of mixing, since, in a grid-based code without explicit diffusion such as ours, the chemical (and physical) diffusivity is directly set by the grid size. We do not expect all quantities in our simulations to converge -- indeed, a number of authors have pointed out that there appears to be no converged solution the problem of computing the mass spectrum of objects produced by gravitational fragmentation of a turbulent medium with an isothermal or sub-isothermal equation of state\cite{martel06a, krumholz14c}. We therefore do not expect things like the mass distribution or number of stars in our simulations to converge. However, we can still check if the amount of chemical mixing is converged, or, more basically, if there is a trend of increasing or decreasing mixing with resolution that we can use to extrapolate.

To assess this question, we have performed runs S3, S4, and 512S1, which have identical physical conditions as run S, but differ in resolution. Runs S, S3, and S4 all have the same base grid resolution, but differ in the maximum AMR level permitted before sink particles are introduced. Since refinement is based on the Jeans condition, these runs are therefore identical in their resolution of low-density, non-self-gravitating gas, but runs S3 and S4 offer factors of 2 and 4, respectively, better resolution in the self-gravitating regions from which stars form. The corresponding minimum cell sizes are $\Delta x_{\rm min} = 1/16$ and $1/32$ pc. In comparison, run 512S1 has the same peak resolution as S ($\Delta x_{\rm min} = 1/8$ pc), but uses twice as many cells in its base grid. Run 512S1 therefore provides better resolution in the diffuse, non-self-gravitating gas ($\Delta x = 1/4$ pc), but the same resolution in self-gravitating regions.

\begin{figure}
\begin{center}
\includegraphics[width=0.5\textwidth]{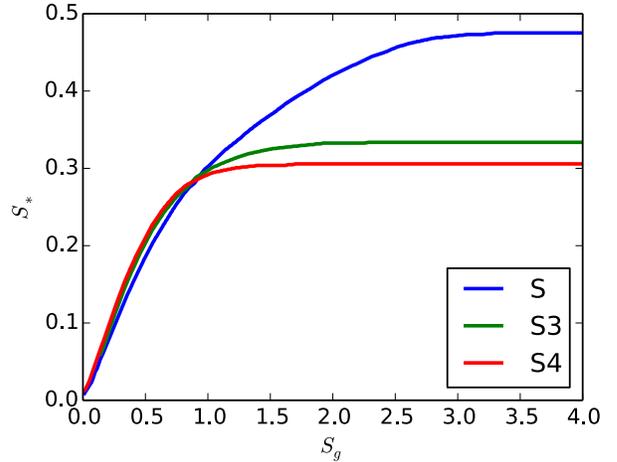}
\end{center}
\caption{
\label{fig:converge}
Stellar abundance scatter $S_{*}$ as a function of gas abundance scatter $S_{g}$ for runs S, S3, and S4, measured at the time when the star formation efficiency $\varepsilon \approx 0.06$.
}
\end{figure}

\begin{figure}
\begin{center}
\includegraphics[width=0.5\textwidth]{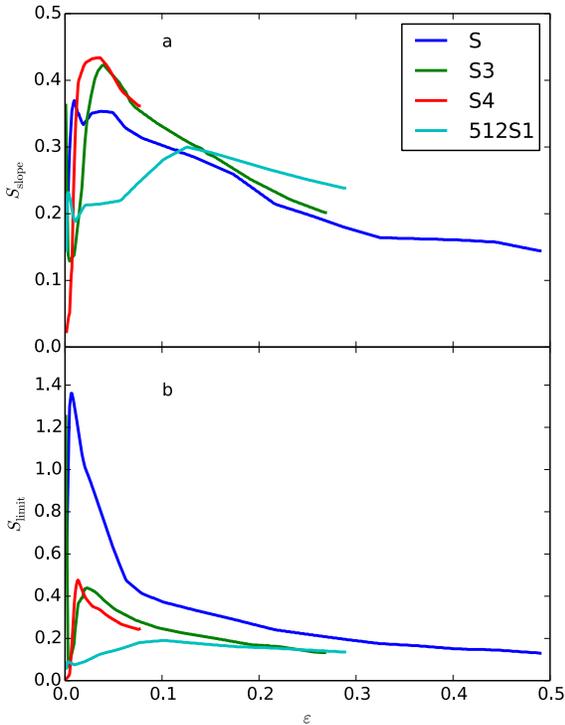} 
\end{center}
\caption{
\label{fig:sslope_limit_converge}
Evolution of two measures of the abundance scatter versus star formation efficiency $\varepsilon$ in runs S, S3, S4, and 512S1. (a) Evolution of $S_{\rm slope}$, the factor by which the abundance scatter is reduced in the limit where the gas abundance scatter $S_g$ is small. (b) Evolution of $S_{\rm limit}$, the maximum stellar abundance scatter in the limit of infinite gas abundance scatter.
}
\end{figure}

We plot $S_{\rm limit}$ and $S_{\rm slope}$ as a function of star formation efficiency $\varepsilon$ for all runs in Figure \ref{fig:converge}. We also plot $S_{*}$ as a function of $S_{g}$ for runs S, S3, and S4 at a fixed star formation efficiency $\varepsilon\approx 0.06$ in Figure \ref{fig:converge}. For the runs that have base grid of 256 cells, the plots show strong signs of convergence. Qualitatively, runs S, S3, and S4 all show similar variations of $S_*$ and $S_{\rm limit}$ versus $\varepsilon$. The only substantial difference is for $S_{\rm limit}$ in run S at very low $\varepsilon$, when there are very few stars present and the results are therefore highly stochastic. In contrast, for runs S3 and S4, $S_{\rm slope}$ values are almost the same even when the star formation efficiency $\varepsilon$ is less than 0.1. The values of $S_{\rm limit}$ in these runs are within 10\% of one another at all times. We can see this even more clearly from Figure \ref{fig:sslope_limit_converge}, where the curves of $S_{*}$ versus $S_{g}$ for runs S3 and S4 are fairly close. The two curves overlap when $S_{g} < 1$ and differ only slightly when $S_{g} \rightarrow \infty$.

In comparison, the time evolution of $S_*$ and $S_{\rm limit}$ in run 512S1 is somewhat qualitatively different: the absolute level of scatter is smaller than in run S at all times, and the value of $S_{\rm limit}$ is smaller than in runs S3 and S4 at almost all times as well, while the value of $S_*$ is about the same. Increasing the base grid resolution therefore also appears to result in reduced scatter. This may occur because, with finer resolution of the base grid, turbulent mixing is better resolved in lower density regions, thus make the mixing process more efficient. So the rapid rise of scatter initially of runs with base grid resolution of 256 may be caused by the gravitational collapse of poorly mixed regions. The same regions are better mixed in run 512S1 because of better resolved turbulence, so the rapid rise disappears in run 512S1 and the scatter is also smaller thereafter.

Thus we find that increasing either the finest AMR level or the number of cells per linear dimension on the coarsest AMR level reduces the abundance scatter, although the results converge relatively fast and especially $S_{\rm slope}$ does not change much between different runs. This result might initially seem surprising, since increasing resolution should decrease the diffusivity of the code. However, this effect appears to be outweighed by the better resolution of the turbulence, perhaps associated with a better separation between the warm and cold phases, provided by a finer grid. In any event, the conclusion that the abundance scatter is reduced by a factor of $\sim 3-6$ compared to that in the gas from which the stars are formed appears to be robust to changes in resolution. If anything, we have underestimated the degree of mixing.\\

\noindent
\textbf{Implications and Broader Context.}\\

Chemical homogeneity in open star clusters is of interest for a number of reasons, and our results therefore have a number of important implications. One important implication is for chemical tagging\cite{freeman02a}. The field stars that make up the bulk of the Galactic disk most likely consist of dissolved star clusters, each carrying a unique chemical tag marking its birth site. If one measures the abundances of enough elements in field stars with enough precision, then in principle it should be possible to use this tag to infer that two seemingly-unrelated field stars in fact originated in the same cluster, and in the process one could answer a number of outstanding questions about the origin of the Sun and the movements of stars through the Galaxy\cite{portegies-zwart09a, bland-hawthorn10a}. There are two major surveys underway that include among their goals performing such reconstructions: the \textit{Gaia}-ESO Public Spectroscopic Survey\cite{gilmore12a}, and the Galactic Archaeology with HERMES Survey.

However, in the absence of a theoretical explanation for why star clusters are chemically homogenous, and under what circumstances we expect homogeneity to prevail, doubt must remain about these techniques. For example, observations indicate that only a small fraction of star formation produces gravitationally-bound open clusters\cite{lada03a}, and it might be the case that the observed chemically-homogenous open clusters and moving groups represent a mode of star formation that produces an unusually-high degree of chemical homogeneity. Our work showing that chemical homogeneity is achieved even at very low star formation efficiencies should help lay this concern to rest. Of course for chemical tagging to be practical it must be the case that clusters are not only internally chemically-homogenous, but that they are also sufficiently distinct from one another that it is possible to distinguish between them using the types of moderate-resolution spectra that can be obtained for large samples\cite{bland-hawthorn10a, bland-hawthorn10b, 2013MNRAS.428.2321M}, and our study of individual clusters does not address this second requirement. However, recent practical successes in using chemical tagging to identify coeval stellar groups\cite{2014MNRAS.438.2753M} suggest that it is satisfied as well.

More broadly, our work has identified a powerful new process for chemically homogenizing the stars in a galactic disk. Observations show that, at a given overall metallicity, stars in the Milky Way thin disk show remarkably little abundance variation\cite{bensby14a}. This must partly be due to homogenization of the interstellar medium (ISM) across galactic scales, which wipe out large-scale variations in chemical abundance. Supernova-driven turbulence appears to mix large-scale modes most effectively\cite{de-avillez02a}, while thermal instability mixes small modes but still leaves small filaments of cold ISM with noticeably different abundances than their surroundings\cite{2012ApJ...758...48Y}. These mechanisms likely explain the relatively low logarithmic abundance scatter of $\sim 0.1$ seen in the ISM. However, to explain the even smaller spread in stellar abundances seen on galactic scales likely requires a further mixing mechanism that can wipe out inhomogeneities on very small scales within star-forming clouds. It seems likely that the turbulent mixing mechanism we have identified in this work is responsible.

\bibliographystyle{naturemag}
\bibliography{version1}

\begin{thebibliography}{10}
\expandafter\ifx\csname url\endcsname\relax
  \def\url#1{\texttt{#1}}\fi
\expandafter\ifx\csname urlprefix\endcsname\relax\def\urlprefix{URL }\fi
\providecommand{\bibinfo}[2]{#2}
\providecommand{\eprint}[2][]{\url{#2}}

\bibitem{2007AJ....133..694D}
\bibinfo{author}{{De Silva}, G.~M.}, \bibinfo{author}{{Freeman}, K.~C.},
  \bibinfo{author}{{Bland-Hawthorn}, J.}, \bibinfo{author}{{Asplund}, M.} \&
  \bibinfo{author}{{Bessell}, M.~S.}
\newblock \bibinfo{title}{{Chemically Tagging the HR 1614 Moving Group}}.
\newblock \emph{\bibinfo{journal}{\aj}} \textbf{\bibinfo{volume}{133}},
  \bibinfo{pages}{694--704} (\bibinfo{year}{2007}).

\bibitem{2007AJ....133.1161D}
\bibinfo{author}{{De Silva}, G.~M.} \emph{et~al.}
\newblock \bibinfo{title}{{Chemical Homogeneity in Collinder 261 and
  Implications for Chemical Tagging}}.
\newblock \emph{\bibinfo{journal}{\aj}} \textbf{\bibinfo{volume}{133}},
  \bibinfo{pages}{1161--1175} (\bibinfo{year}{2007}).

\bibitem{pancino10a}
\bibinfo{author}{{Pancino}, E.}, \bibinfo{author}{{Carrera}, R.},
  \bibinfo{author}{{Rossetti}, E.} \& \bibinfo{author}{{Gallart}, C.}
\newblock \bibinfo{title}{{Chemical abundance analysis of the open clusters Cr
  110, NGC 2099 (M 37), NGC 2420, NGC 7789, and M 67 (NGC 2682)}}.
\newblock \emph{\bibinfo{journal}{\aap}} \textbf{\bibinfo{volume}{511}},
  \bibinfo{pages}{A56} (\bibinfo{year}{2010}).
\newblock \eprint{0910.0723}.

\bibitem{bubar10a}
\bibinfo{author}{{Bubar}, E.~J.} \& \bibinfo{author}{{King}, J.~R.}
\newblock \bibinfo{title}{{Spectroscopic Abundances and Membership in the Wolf
  630 Moving Group}}.
\newblock \emph{\bibinfo{journal}{\aj}} \textbf{\bibinfo{volume}{140}},
  \bibinfo{pages}{293--318} (\bibinfo{year}{2010}).

\bibitem{de-silva11a}
\bibinfo{author}{{de Silva}, G.~M.} \emph{et~al.}
\newblock \bibinfo{title}{{High-resolution elemental abundance analysis of the
  Hyades supercluster}}.
\newblock \emph{\bibinfo{journal}{\mnras}} \textbf{\bibinfo{volume}{415}},
  \bibinfo{pages}{563--575} (\bibinfo{year}{2011}).

\bibitem{ting12a}
\bibinfo{author}{{Ting}, Y.-S.}, \bibinfo{author}{{De Silva}, G.~M.},
  \bibinfo{author}{{Freeman}, K.~C.} \& \bibinfo{author}{{Parker}, S.~J.}
\newblock \bibinfo{title}{{High-resolution elemental abundance analysis of the
  open cluster IC 4756}}.
\newblock \emph{\bibinfo{journal}{\mnras}} \textbf{\bibinfo{volume}{427}},
  \bibinfo{pages}{882--892} (\bibinfo{year}{2012}).

\bibitem{reddy12b}
\bibinfo{author}{{Reddy}, A.~B.~S.}, \bibinfo{author}{{Giridhar}, S.} \&
  \bibinfo{author}{{Lambert}, D.~L.}
\newblock \bibinfo{title}{{Comprehensive abundance analysis of red giants in
  the open clusters NGC 752, 1817, 2360 and 2506}}.
\newblock \emph{\bibinfo{journal}{\mnras}} \textbf{\bibinfo{volume}{419}},
  \bibinfo{pages}{1350--1361} (\bibinfo{year}{2012}).

\bibitem{de-silva13a}
\bibinfo{author}{{De Silva}, G.~M.} \emph{et~al.}
\newblock \bibinfo{title}{{Search for associations containing young stars:
  chemical tagging IC 2391 and the Argus association}}.
\newblock \emph{\bibinfo{journal}{\mnras}} \textbf{\bibinfo{volume}{431}},
  \bibinfo{pages}{1005--1018} (\bibinfo{year}{2013}).

\bibitem{reddy13a}
\bibinfo{author}{{Reddy}, A.~B.~S.}, \bibinfo{author}{{Giridhar}, S.} \&
  \bibinfo{author}{{Lambert}, D.~L.}
\newblock \bibinfo{title}{{Comprehensive abundance analysis of red giants in
  the open clusters NGC 2527, 2682, 2482, 2539, 2335, 2251 and 2266}}.
\newblock \emph{\bibinfo{journal}{\mnras}} \textbf{\bibinfo{volume}{431}},
  \bibinfo{pages}{3338--3348} (\bibinfo{year}{2013}).

\bibitem{2008ApJ...675.1213R}
\bibinfo{author}{{Rosolowsky}, E.} \& \bibinfo{author}{{Simon}, J.~D.}
\newblock \bibinfo{title}{{The M33 Metallicity Project: Resolving the Abundance
  Gradient Discrepancies in M33}}.
\newblock \emph{\bibinfo{journal}{\apj}} \textbf{\bibinfo{volume}{675}},
  \bibinfo{pages}{1213--1222} (\bibinfo{year}{2008}).

\bibitem{2012ApJ...758..133S}
\bibinfo{author}{{Sanders}, N.~E.}, \bibinfo{author}{{Caldwell}, N.},
  \bibinfo{author}{{McDowell}, J.} \& \bibinfo{author}{{Harding}, P.}
\newblock \bibinfo{title}{{The Metallicity Profile of M31 from Spectroscopy of
  Hundreds of H II Regions and PNe}}.
\newblock \emph{\bibinfo{journal}{\apj}} \textbf{\bibinfo{volume}{758}},
  \bibinfo{pages}{133} (\bibinfo{year}{2012}).

\bibitem{2013ApJ...775..128B}
\bibinfo{author}{{Berg}, D.~A.} \emph{et~al.}
\newblock \bibinfo{title}{{New Radial Abundance Gradients for NGC 628 and NGC
  2403}}.
\newblock \emph{\bibinfo{journal}{\apj}} \textbf{\bibinfo{volume}{775}},
  \bibinfo{pages}{128} (\bibinfo{year}{2013}).

\bibitem{2011ApJ...730..129B}
\bibinfo{author}{{Bresolin}, F.}
\newblock \bibinfo{title}{{The Abundance Scatter in M33 from H II Regions: Is
  There Any Evidence for Azimuthal Metallicity Variations?}}
\newblock \emph{\bibinfo{journal}{\apj}} \textbf{\bibinfo{volume}{730}},
  \bibinfo{pages}{129} (\bibinfo{year}{2011}).

\bibitem{2013ApJ...766...17L}
\bibinfo{author}{{Li}, Y.}, \bibinfo{author}{{Bresolin}, F.} \&
  \bibinfo{author}{{Kennicutt}, R.~C., Jr.}
\newblock \bibinfo{title}{{Testing for Azimuthal Abundance Gradients in M101}}.
\newblock \emph{\bibinfo{journal}{\apj}} \textbf{\bibinfo{volume}{766}},
  \bibinfo{pages}{17} (\bibinfo{year}{2013}).

\bibitem{carroll-nellenback13a}
\bibinfo{author}{{Carroll-Nellenback}, J.}, \bibinfo{author}{{Frank}, A.} \&
  \bibinfo{author}{{Heitsch}, F.}
\newblock \bibinfo{title}{{The Effects of Inhomogeneities within Colliding
  Flows on the Formation and Evolution of Molecular Clouds}}.
\newblock \emph{\bibinfo{journal}{\apj}}  (\bibinfo{year}{2013}).
\newblock \bibinfo{note}{Submitted, arXiv:1034.1367}.

\bibitem{murray90a}
\bibinfo{author}{{Murray}, S.~D.} \& \bibinfo{author}{{Lin}, D.~N.~C.}
\newblock \bibinfo{title}{{On the origin of metal homogeneities in globular
  clusters}}.
\newblock \emph{\bibinfo{journal}{\apj}} \textbf{\bibinfo{volume}{357}},
  \bibinfo{pages}{105--112} (\bibinfo{year}{1990}).

\bibitem{de-avillez02a}
\bibinfo{author}{{de Avillez}, M.~A.} \& \bibinfo{author}{{Mac Low}, M.-M.}
\newblock \bibinfo{title}{{Mixing Timescales in a Supernova-driven Interstellar
  Medium}}.
\newblock \emph{\bibinfo{journal}{\apj}} \textbf{\bibinfo{volume}{581}},
  \bibinfo{pages}{1047--1060} (\bibinfo{year}{2002}).

\bibitem{2012ApJ...758...48Y}
\bibinfo{author}{{Yang}, C.-C.} \& \bibinfo{author}{{Krumholz}, M.}
\newblock \bibinfo{title}{{Thermal-instability-driven Turbulent Mixing in
  Galactic Disks. I. Effective Mixing of Metals}}.
\newblock \emph{\bibinfo{journal}{\apj}} \textbf{\bibinfo{volume}{758}},
  \bibinfo{pages}{48} (\bibinfo{year}{2012}).

\bibitem{1998ApJ...495..821T}
\bibinfo{author}{{Truelove}, J.~K.} \emph{et~al.}
\newblock \bibinfo{title}{{Self-gravitational Hydrodynamics with
  Three-dimensional Adaptive Mesh Refinement: Methodology and Applications to
  Molecular Cloud Collapse and Fragmentation}}.
\newblock \emph{\bibinfo{journal}{\apj}} \textbf{\bibinfo{volume}{495}},
  \bibinfo{pages}{821--852} (\bibinfo{year}{1998}).

\bibitem{Klein99a}
\bibinfo{author}{{Klein}, R.~I.}
\newblock \bibinfo{title}{{Star formation with 3-D adaptive mesh refinement:
  the collapse and fragmentation of molecular clouds}}.
\newblock \emph{\bibinfo{journal}{JCAM}} \textbf{\bibinfo{volume}{109}},
  \bibinfo{pages}{123--152} (\bibinfo{year}{1999}).

\bibitem{2004ApJ...611..399K}
\bibinfo{author}{{Krumholz}, M.~R.}, \bibinfo{author}{{McKee}, C.~F.} \&
  \bibinfo{author}{{Klein}, R.~I.}
\newblock \bibinfo{title}{{Embedding Lagrangian Sink Particles in Eulerian
  Grids}}.
\newblock \emph{\bibinfo{journal}{\apj}} \textbf{\bibinfo{volume}{611}},
  \bibinfo{pages}{399--412} (\bibinfo{year}{2004}).

\bibitem{2007ApJ...657..870V}
\bibinfo{author}{{V{\'a}zquez-Semadeni}, E.} \emph{et~al.}
\newblock \bibinfo{title}{{Molecular Cloud Evolution. II. From Cloud Formation
  to the Early Stages of Star Formation in Decaying Conditions}}.
\newblock \emph{\bibinfo{journal}{\apj}} \textbf{\bibinfo{volume}{657}},
  \bibinfo{pages}{870--883} (\bibinfo{year}{2007}).

\bibitem{2008ApJ...674..316H}
\bibinfo{author}{{Heitsch}, F.}, \bibinfo{author}{{Hartmann}, L.~W.},
  \bibinfo{author}{{Slyz}, A.~D.}, \bibinfo{author}{{Devriendt}, J.~E.~G.} \&
  \bibinfo{author}{{Burkert}, A.}
\newblock \bibinfo{title}{{Cooling, Gravity, and Geometry: Flow-driven Massive
  Core Formation}}.
\newblock \emph{\bibinfo{journal}{\apj}} \textbf{\bibinfo{volume}{674}},
  \bibinfo{pages}{316--328} (\bibinfo{year}{2008}).

\bibitem{tan06a}
\bibinfo{author}{{Tan}, J.~C.}, \bibinfo{author}{{Krumholz}, M.~R.} \&
  \bibinfo{author}{{McKee}, C.~F.}
\newblock \bibinfo{title}{{Equilibrium Star Cluster Formation}}.
\newblock \emph{\bibinfo{journal}{\apjl}} \textbf{\bibinfo{volume}{641}},
  \bibinfo{pages}{L121--L124} (\bibinfo{year}{2006}).

\bibitem{krumholz07e}
\bibinfo{author}{{Krumholz}, M.~R.} \& \bibinfo{author}{{Tan}, J.~C.}
\newblock \bibinfo{title}{{Slow Star Formation in Dense Gas: Evidence and
  Implications}}.
\newblock \emph{\bibinfo{journal}{\apj}} \textbf{\bibinfo{volume}{654}},
  \bibinfo{pages}{304--315} (\bibinfo{year}{2007}).

\bibitem{krumholz14c}
\bibinfo{author}{{Krumholz}, M.~R.}
\newblock \bibinfo{title}{{The Big Problems in Star Formation: the Star
  Formation Rate, Stellar Clustering, and the Initial Mass Function}}.
\newblock \emph{\bibinfo{journal}{Physics Reports}}
  \textbf{\bibinfo{volume}{539}}, \bibinfo{pages}{49--134}
  (\bibinfo{year}{2014}).

\bibitem{portegies-zwart09a}
\bibinfo{author}{{Portegies Zwart}, S.~F.}
\newblock \bibinfo{title}{{The Lost Siblings of the Sun}}.
\newblock \emph{\bibinfo{journal}{\apjl}} \textbf{\bibinfo{volume}{696}},
  \bibinfo{pages}{L13--L16} (\bibinfo{year}{2009}).

\bibitem{bland-hawthorn10a}
\bibinfo{author}{{Bland-Hawthorn}, J.}, \bibinfo{author}{{Krumholz}, M.~R.} \&
  \bibinfo{author}{{Freeman}, K.}
\newblock \bibinfo{title}{{The Long-term Evolution of the Galactic Disk Traced
  by Dissolving Star Clusters}}.
\newblock \emph{\bibinfo{journal}{\apj}} \textbf{\bibinfo{volume}{713}},
  \bibinfo{pages}{166--179} (\bibinfo{year}{2010}).

\bibitem{2013MNRAS.428.2321M}
\bibinfo{author}{{Mitschang}, A.~W.}, \bibinfo{author}{{De Silva}, G.},
  \bibinfo{author}{{Sharma}, S.} \& \bibinfo{author}{{Zucker}, D.~B.}
\newblock \bibinfo{title}{{Quantifying chemical tagging: towards robust group
  finding in the Galaxy}}.
\newblock \emph{\bibinfo{journal}{\mnras}} \textbf{\bibinfo{volume}{428}},
  \bibinfo{pages}{2321--2332} (\bibinfo{year}{2013}).

\bibitem{2014MNRAS.438.2753M}
\bibinfo{author}{{Mitschang}, A.~W.} \emph{et~al.}
\newblock \bibinfo{title}{{Quantitative chemical tagging, stellar ages and the
  chemo-dynamical evolution of the Galactic disc}}.
\newblock \emph{\bibinfo{journal}{\mnras}} \textbf{\bibinfo{volume}{438}},
  \bibinfo{pages}{2753--2764} (\bibinfo{year}{2014}).

\bibitem{2002ApJ...564L..97K}
\bibinfo{author}{{Koyama}, H.} \& \bibinfo{author}{{Inutsuka}, S.-i.}
\newblock \bibinfo{title}{{An Origin of Supersonic Motions in Interstellar
  Clouds}}.
\newblock \emph{\bibinfo{journal}{\apj}} \textbf{\bibinfo{volume}{564}},
  \bibinfo{pages}{L97--L100} (\bibinfo{year}{2002}).

\bibitem{hockney88a}
\bibinfo{author}{{Hockney}, R.~W.} \& \bibinfo{author}{{Eastwood}, J.~W.}
\newblock \emph{\bibinfo{title}{{Computer simulation using particles}}}
  (\bibinfo{publisher}{CRC Press}, \bibinfo{year}{1988}).

\bibitem{martel06a}
\bibinfo{author}{{Martel}, H.}, \bibinfo{author}{{Evans}, N.~J., II} \&
  \bibinfo{author}{{Shapiro}, P.~R.}
\newblock \bibinfo{title}{{Fragmentation and Evolution of Molecular Clouds. I.
  Algorithm and First Results}}.
\newblock \emph{\bibinfo{journal}{\apjs}} \textbf{\bibinfo{volume}{163}},
  \bibinfo{pages}{122--144} (\bibinfo{year}{2006}).

\bibitem{freeman02a}
\bibinfo{author}{{Freeman}, K.} \& \bibinfo{author}{{Bland-Hawthorn}, J.}
\newblock \bibinfo{title}{{The New Galaxy: Signatures of Its Formation}}.
\newblock \emph{\bibinfo{journal}{\araa}} \textbf{\bibinfo{volume}{40}},
  \bibinfo{pages}{487--537} (\bibinfo{year}{2002}).

\bibitem{gilmore12a}
\bibinfo{author}{{Gilmore}, G.} \emph{et~al.}
\newblock \bibinfo{title}{{The Gaia-ESO Public Spectroscopic Survey}}.
\newblock \emph{\bibinfo{journal}{The Messenger}}
  \textbf{\bibinfo{volume}{147}}, \bibinfo{pages}{25--31}
  (\bibinfo{year}{2012}).

\bibitem{lada03a}
\bibinfo{author}{{Lada}, C.~J.} \& \bibinfo{author}{{Lada}, E.~A.}
\newblock \bibinfo{title}{{Embedded Clusters in Molecular Clouds}}.
\newblock \emph{\bibinfo{journal}{\araa}} \textbf{\bibinfo{volume}{41}},
  \bibinfo{pages}{57--115} (\bibinfo{year}{2003}).

\bibitem{bland-hawthorn10b}
\bibinfo{author}{{Bland-Hawthorn}, J.}, \bibinfo{author}{{Karlsson}, T.},
  \bibinfo{author}{{Sharma}, S.}, \bibinfo{author}{{Krumholz}, M.} \&
  \bibinfo{author}{{Silk}, J.}
\newblock \bibinfo{title}{{The Chemical Signatures of the First Star Clusters
  in the Universe}}.
\newblock \emph{\bibinfo{journal}{\apj}} \textbf{\bibinfo{volume}{721}},
  \bibinfo{pages}{582--596} (\bibinfo{year}{2010}).

\bibitem{bensby14a}
\bibinfo{author}{{Bensby}, T.}, \bibinfo{author}{{Feltzing}, S.} \&
  \bibinfo{author}{{Oey}, M.~S.}
\newblock \bibinfo{title}{{Exploring the Milky Way stellar disk. A detailed
  elemental abundance study of 714 F and G dwarf stars in the solar
  neighbourhood}}.
\newblock \emph{\bibinfo{journal}{\aap}} \textbf{\bibinfo{volume}{562}},
  \bibinfo{pages}{A71} (\bibinfo{year}{2014}).
\newblock \eprint{1309.2631}.

\end{thebibliography}

\end{document}